# Coulomb-interaction induced incomplete shell filling in the hole system of InAs quantum dots

D. Reuter[1,2], P. Kailuweit[1], A. D. Wieck[1], U. Zeitler[2], O. S. Wibbelhoff[3], C. Meier[3], A. Lorke[3], and J. C. Maan[2]

[1]Lehrstuhl für Angewandte Festkörperphysik, Ruhr-Universität Bochum, D-44799 Bochum, Germany

[2]High Field Magnet Laboratory, NSRIM, University of Nijmegen, Toernooiveld 7, 6525 ED Nijmegen, The Netherlands

[3]Experimentalphysik, Universität Duisburg-Essen, Lotharstraße 1, D-47048 Duisburg, Germany

We have studied the hole charging spectra of self-assembled InAs quantum dots in perpendicular magnetic fields by capacitance-voltage spectroscopy. From the magnetic field dependence of the individual peaks we conclude that the $s$-like ground state is completely filled with two holes but that the fourfold degenerate $p$-shell is only half filled with two holes before the filling of the $d$-shell starts. The resulting six-hole ground state is highly polarized. This incomplete shell filling can be explained by the large influence of the Coulomb interaction in this system.

73.21.La, 73.22.-f, 73.23.Hk, 73.63.Kv

Correlation effects due to Coulomb interaction are fundamental for the understanding of numerous physical phenomena such as magnetism, superconductivity and collective ground states in the fractional quantum Hall effect. In general, correlations are visible in the collective behavior of a great number of charged particles and the relevant energies are therefore hard to access directly. However, in semiconductor quantum dots (QDs) occupied only by a few carriers, such a direct access becomes possible. In these systems, often regarded as artificial atoms due to their shell-like energy spectrum, the correlation energies are very important in determining the charging sequence of the individual shells and can be directly accessed as states are filled one by one with particles[1,2].

Semiconductor QDs can be fabricated by many methods, e. g. by lithography[2] or by self-organized growth[1]. As one of the most prominent realizations of three-dimensional carrier confinement, self-assembled InAs (or $In_xGa_{1-x}As$) QDs have been studied in great detail[1]. Beside their importance for basic research on zero-dimensional carrier systems, these systems might also have technological applications, e. g., in low threshold lasers[1], single photon sources[3], or solid-state implementations of quantum bits. On a more fundamental level, one of the main research goals is to reveal their precise level structure including both confinement effects and Coulomb interactions.

Up until the present, mainly conduction band states have been studied in this context where capacitance-voltage (*C-V*) spectroscopy provided valuable information on the level structure and the Coulomb interactions[4,5,6]. A non-sequential level filling for degenerate or nearly degenerate, respectively, energy levels, i. e., the manifestation of Hund´s first rule, was observed in some experiments on self-assembled InAs QDs[7]. This behavior was also observed in lithographically defined QDs, based on a GaAs/Al$_x$Ga$_{1-x}$As resonant tunneling structure[2]. For both systems, the results could be well explained assuming a lateral confinement by a harmonic oscillator potential and treating the Coulomb interaction as a perturbation[2,6]. As required for perturbation theory, for electrons in InAs QDs, the Coulomb energies are significantly (two to three times) smaller than the quantization energy and this approach reproduces the data quantitatively[6]. Also for lithographically defined QDs, the quantitative agreement is quite good[6] although Coulomb energies and quantization energy are about the same. In neither of the two types of QDs incomplete shell filling with respect to non-degenerate single particle levels, as will be discussed in this Letter, was observed.

Due to the larger effective mass, the spatial carrier confinement in hole systems is stronger than in electron systems which results in a stronger Coulomb interaction and smaller the quantization energies This different ratio between these two important energy scales opens the possibility for novel correlation phenomena in a few carrier QD. Holes in InAs QDs are much less investigated[8] than electrons and only very recently high quality *C-V* measurements became available[9,10] making investigations of the hole addition spectra feasible.

In this Letter, we present experimental results on the shell filling of holes in self-assembled InAs QDs, a system where the direct part as well as the exchange contribution of the Coulomb interaction are very similar to the quantization energy. *C-V* measurements in perpendicular magnetic fields up to 32 T directly show the peculiar nature of the hole levels and demonstrates that the strong influence of Coulomb interactions even affects the level filling sequence. The observed non-sequential level filling caused by Coulomb interactions, which results in incompletely filled shells corresponding to a highly polarized few hole ground state.

The InAs QD samples were prepared by solid source molecular beam epitaxy on a GaAs(100) substrate. The active part of the layer sequence consists of a 300 nm thick, carbon doped ($3\times10^{18}$ cm$^{-3}$) GaAs back-contact, a 17 nm thick GaAs tunneling barrier, an InAs QD layer, 30 nm GaAs and 27 periods of a (3 nm AlAs)/(1 nm GaAs) superlattice, followed by a 10 nm thick GaAs cap layer. The InAs QDs were prepared by depositing a nominal coverage of 2.0 ML InAs at a substrate temperature of 510 °C. The ground state photoluminescence for these samples is between 1250 and 1270 nm at 300 K. From these samples, Schottky diodes were prepared using Cr-Au gates. The *C-V* traces were recorded at 4.2 K with a standard LCR meter (Agilent 4284A).

In Fig. 1, *C-V* spectra of the hole system for various magnetic fields *B* (oriented perpendicular to the base of the InAs QDs, i.e., along the growth direction) are shown. For $B = 0$ T, six clear charging peaks and a broader one, probably composed of two peaks, can be identified. By fitting peaks 1-6 with Gaussians, we can well reproduce the spectra. The areas of all peaks agree within 20 % and the charge obtained from the peak area agrees well with a dot density of ~$2\times10^{10}$ cm$^{-2}$ expected for this growth condition. Therefore, we conclude that each of the peaks corresponds to the charging of only one additional hole per QD. The peaks 1 and 2 exhibit almost no shift with magnetic field whereas the peaks 3 to 6 show – partly non-monotonic - shifts with magnetic field.

In Fig. 2, the energetic positions of the individual charging peaks as a function of the magnetic field are shown. The gate voltage scale was converted to an energy scale, taking the voltage-dependent depletion in the back gate into account (see Ref. 9). We find that the hole ground state is situated 204 meV below the GaAs valence band edge. The direct Coulomb energy (separation between peaks 1 and 2) for the ground state is 24 meV. These values agree well with the observations of Bock and co-workers on QDs showing similar photoluminescence spectra[10]. For $B < 14$ T, peaks 3 and 5 shift downwards and peaks 4 and 6 shift upwards in energy. Above B = 16 T the shifting direction is reversed for the peaks 4, 5 and 6.

From the change in the energy difference between peaks 1 and 2 (using the formula $E_z = 2\mu_b g_h B |J_z|$, where $\mu_b$ is the Bohr magneton and $J_z = \pm 3/2$ as discussed below), we estimate an upper limit of 0.5 for the hole g-factor $g_h$ in our QDs. Therefore, it is clear that the shifts observed for peaks 3 to 6 are not caused by a hole spin Zeeman effect, i. e., the interaction of the magnetic field with the Bloch character of the hole. Consequently, we attribute the observed shifts to the orbital Zeeman effect. This is further supported by *C-V* experiments in parallel magnetic fields, i.e., the field is applied inside the base plane of the InAs QDs. Up to 28 T, the observed shifts were not exceeding 1.5 meV for any charging peak and were, for most peaks, below the experimental resolution. This is consistent with an angular momentum $L_z$ perpendicular to the base plane of the quantum dots, i. e., a movement of the carriers parallel to the base plane, as expected for a strong anisotropy in the spatial extension of the carrier wave function[11].

To demonstrate the different slopes of the individual charging peaks in the low field range ($B \leq 14$ T), Fig. 3 shows the absolute value of the changes in peak position with magnetic field, i. e., the energy for $B = 0$ was subtracted. One can clearly distinguish three different slopes: Peaks 1 and 2 exhibit almost no shift. Peaks 3 and 4 shift by (0.17±0.02 meV/T) and (0.14±0.02 meV/T), respectively, while charging maxima 5 (0.35±0.02 meV/T) and 6 (0.33±0.02 meV/T) show a magnetic field dependence that is approximately twice as strong as for peaks 3 and 4.

In the following, we will discuss in more detail the nature of the individual charging peaks. In any rotationally symmetric confining potential, especially the two-dimensional harmonic oscillator, which is commonly used to describe the electron system in InAs QDs[6], one obtains single particle levels that can be classified according to their angular momentum as $s$-, $p$- and $d$-states in analogy to atomic physics. To first order, $s$-states do not shift with magnetic field, whereas $d$-states shift twice as strongly as $p$-states (assuming a constant effective mass as discussed below). Therefore, we identify peaks 1 and 2 as $s$-like states, 3 and 4 as $p$-states and 5 and 6 as $d$-like states. Taking the sign of the slopes into account, the six-hole ground state for $0 \leq B \leq 14$ T is then ($ssp_-p_+d_-d_+$). Following the notation of [6], the subscript denotes the direction of the angular momentum $L_z$ states, i.e., $L_z = \pm 1$ for $p$-states, and $L_z = \pm 2$ for $d$-states, respectively. With this assignment we can conclude that, instead of filling the $p$-shell completely with four holes, the filling of the $d$-shell already starts when the $p$-shell is only half filled. The resulting six-hole ground state is almost fully polarized and five out of six holes have $J_z = -3/2$ (see Fig. 2 for a sketch of the configuration). It seems that minimizing the total Coulomb energy, i. e., the gain in exchange energy and a reduction in direct Coulomb energy, not only makes the non-sequential filling of degenerate or nearly-degenerate single particle levels (Hund's rules), e.g., $p_-$ and $p_+$, favorable but can even overcome the energy difference between the $p$- and $d$-shell and lead to the filling of the $d$-shell before the $p$-shell is completed. This is a quite unique feature because for the electron systems in InAs QDs[5,6] or lithographically defined QDs[2] the *p*-shell is filled completely and the six electron ground state (*sspppp*) is unpolarized at $B = 0$ T.

For $B > 16$ T, the magnetic-field-induced energy shifts of the single particle levels lead to a new six-hole ground state: (*ssp₋d₋p₊d₋*), which is reflected in Fig. 2 by the changes in the slope for peaks 4, 5, and 6 around 16 T and can be interpreted as a crossing of single particle levels at this magnetic field.

Because for our peak assignment we rely on the classification of the single particle levels according to angular momentum, it is worth stressing that this assignment is not limited to the harmonic oscillator potential but is valid for any potential for which the angular momentum $L_z$ is a good quantum number. Therefore, any rotationally symmetric potential can be described this way and –in good approximation– also potentials where the dot shape[12] or the ZnS-structure of the material[13] leads to a slight asymmetry. Thus, the discussion of the non-sequential, incomplete level filling, leading to a highly polarized ground state due to important Coulomb interaction is valid even if the exact confinement potential deviates slightly from that of the perfect harmonic oscillator, as long as a single effective mass can be assigned to all relevant single particle levels, which is reasonable for our system as discussed in the next paragraph.

For a more detailed analysis we have to consider that holes are more complicated than electrons because of the complex valence band in III-V semiconductors. In InAs QDs, the degeneracy of the heavy hole ($J_z = \pm 3/2$) and light hole ($J_z = \pm 1/2$) band is lifted due to the carrier confinement. For the lowest lying hole states, a predominantly heavy-hole character ($J_z = \pm 3/2$) with respect to the growth direction is expected because of the significantly smaller quantization energy compared to the light-hole state in growth direction. This was predicted for QDs similar to those discussed here using a simple adiabatic approach[14,15]. The heavy-hole character in growth direction was also confirmed experimentally by optical absorption measurements[14,15]. Furthermore, detailed pseudopotential calculations show the predominantly $J_z = \pm 3/2$ character for the lowest hole states in $In_xGa_{1-x}As$-QDs[13]. Therefore, it is well justified to assume in our case a heavy hole character with respect to growth direction for all observed hole levels and assign to them a single effective mass tensor as done in our discussion above. It is important to stress that due to the anisotropic confinement, i. e., the much stronger confinement in growth direction, for a heavy hole state with respect to motion in growth direction a effective mass closer to the bulk light hole value is expected for motion perpendicular to the growth direction. Note that the effective hole mass for motion perpendicular to the growth direction determines the magnetic field dispersion of the individual charging peaks.

In the following we will compare our data with a widely used model based on a harmonic oscillator potential and perturbation theory, following the procedure outlined in Ref. 6. With this approach, one calculates as criterion that $ssp_-p_+d_-d_+$ becomes the six-hole ground state that the ratio of the quantization energy and the direct Coulomb energy in the $s$-state has to be smaller than 11/16. Taking our measured direct Coulomb energy of 24 meV, the quantization energy should be smaller than 17 meV, which is ~40 % less than the value (29 meV) obtained from the distance between peak 2 and 3 using the same model. Also, the predicted and measured charging energies for peaks 4-6 disagree by about 30 % (5 to 10 meV). This means that the perturbation approach outlined in Ref. 6 describes the system studied here qualitatively, i. e., predicting a $ssp_-p_+d_-d_+$ configuration as six hole ground state if the quantization energy decreases compared to the Coulomb interactions. However, a quantitative description with this approach is accurate only on the 30 % level. On the one hand, this is not surprising because a perturbation approach should not be justified in the case when Coulomb energies and quantization energies are almost equal. On the other hand, it is worth noting that the quantitative agreement is better for electrons in lithographically defined QDs[2,6], where a $ssp_-p_+p_-p_+$ six-electron ground state is observed even though the quantization energy and the direct Coulomb energy in the ground state are roughly the same. This difference might be due to the fact that for our system the balance is shifted even slightly more in favor of the Coulomb interactions. Another reason might be that the spatial confinement is much stronger in our system and, therefore, the error due to the perturbation approach is larger. We believe that only a more elaborate theory can give a complete quantitative description of these results.

To describe the dispersion of the individual charging peaks over the whole magnetic field range we have to go beyond the linear approximation of the orbital Zeeman effect, which requires some knowledge of the specific shape of the confining potential. Lacking a more sophisticated theoretical description, we again use a model of a two-dimensional harmonic oscillator in a perpendicular magnetic field[16,6]. Because a quantitative description of the peak energies is not possible as discussed above, we treat the peak positions at $B = 0\ T$ as fit parameters. With this approach we can describe our data well with a single effective hole mass of $m_h^* = (0.31 \pm 0.03) m_e$, as shown in Fig. 2, using a confinement energy of $\hbar\omega_0 = 25$ meV for the holes estimated from photoluminescence data (not shown). However, it turns out that the exact value of the confinement energy is not very critical for the quality of the fit. We have obtained satisfactory fits for values of $\hbar\omega_0$ between 20 and 30 meV with the corresponding effective masses between 0.3 and $0.32m_e$. The effective hole mass agrees quite well with the value of $0.25m_e$ estimated by Warburton and co-workers[6] from photoluminescence data. The effective mass is significantly smaller than the heavy-hole mass for InAs ($0.41m_e$) and for GaAs ($0.47m_e$)[17]. This is expected because in a layered structure a heavy hole state in growth direction should have light hole character perpendicular to the growth direction[13,14]. However, the effective mass we determine is significantly larger than the values for the light hole in both InAs (*0.025$m_e$*) and GaAs (*0.087$m_e$*). The reason for this large deviation is not completely understood at the moment but is probably related to the strain in the QDs, which is expected to increase the value for the in-plane effective mass. Also non-parabolicity and valence band mixing effects could lead to an increased in-plane hole mass.

In summary, our C-V measurements on the hole system in InAs QDs in magnetic field show a pronounced non-sequential shell filling, a behavior not observed for the corresponding electron system. This incomplete level occupation results in a highly polarized six-hole ground state and shows that the Coulomb interaction can not only enforce the non-sequential filling of degenerate or nearly degenerate single-particle levels but can also overcome the separation between levels associated with different single-particle shells. This is a manifestation of the important role Coulomb interaction plays for the hole system in the InAs QDs investigated.

We gratefully acknowledge financial support by the FOM, the DFG and the BMBF. Part of this work was supported by the European Union (contract no. HPRI-CT-1999-00036).

Figure captions:

Figure 1: Capacitance-voltage spectra for the hole system in InAs quantum dots for various magnetic fields. The traces are off-set vertically for clarity and the lines are guides to the eye to indicate the shifts with magnetic field.

Figure 2: The energetic position of the charging peaks (symbols) from Fig. 1 as function of the magnetic field. The peaks are identified by the same labels as in Fig. 1. N is the total number of holes per QD. The nature of the states with respect to their angular momentum is indicated at the individual graphs. The inserts show the occupation of the single-particle levels for each charging peak before and after the level crossing. The arrows represent $J_z = 3/2$ and $J_z = -3/2$, respectively. The assignment of the squares to different angular momentum states is given in the upper left corner of the figure. The solid lines represent fits, using a harmonic oscillator model as outlined in the text.

Figure 3: The absolute value of the change in energy |E(B)-E(0T)| for charging peaks 1 to 6 (labels according to Fig. 1), for fields up to 15 T. Three classes of slopes can clearly be distinguished, corresponding to different angular momenta: Peaks 1 and 2 show almost no shift ($L_z = 0$), whereas peaks 5 and 6 have twice the slope ($L_z = \pm 2$) as 3 and 4 ($L_z = \pm 1$).

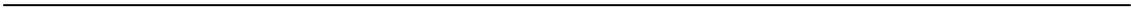

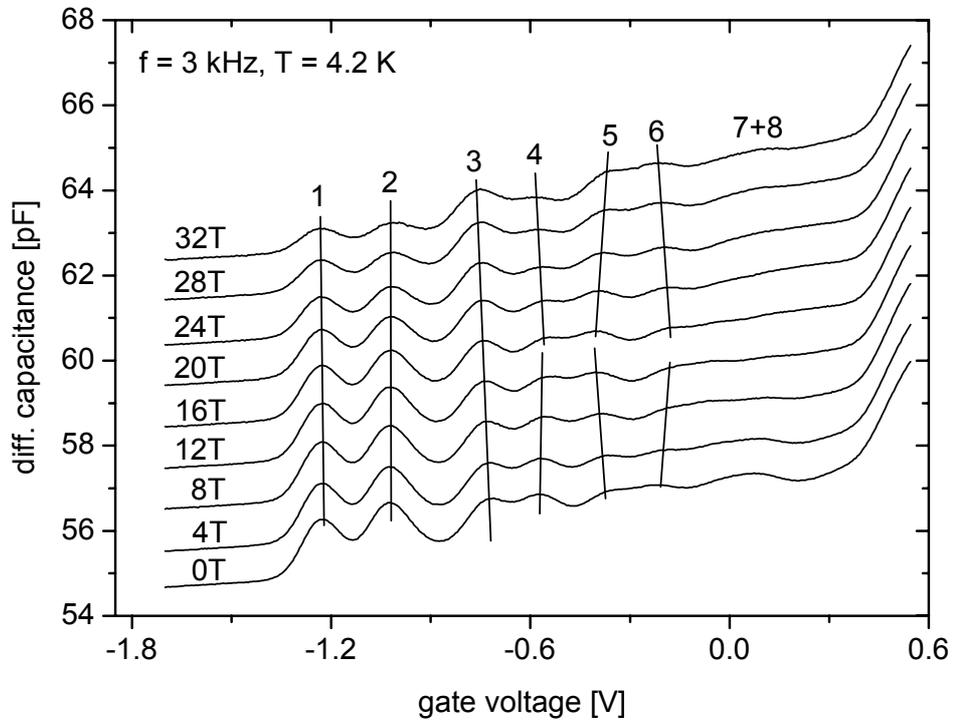

Figure 1: Capacitance-voltage spectra for the hole system in InAs quantum dots for various magnetic fields. The traces are off-set vertically for clarity and the lines are guides to the eye to indicate the shifts with magnetic field.

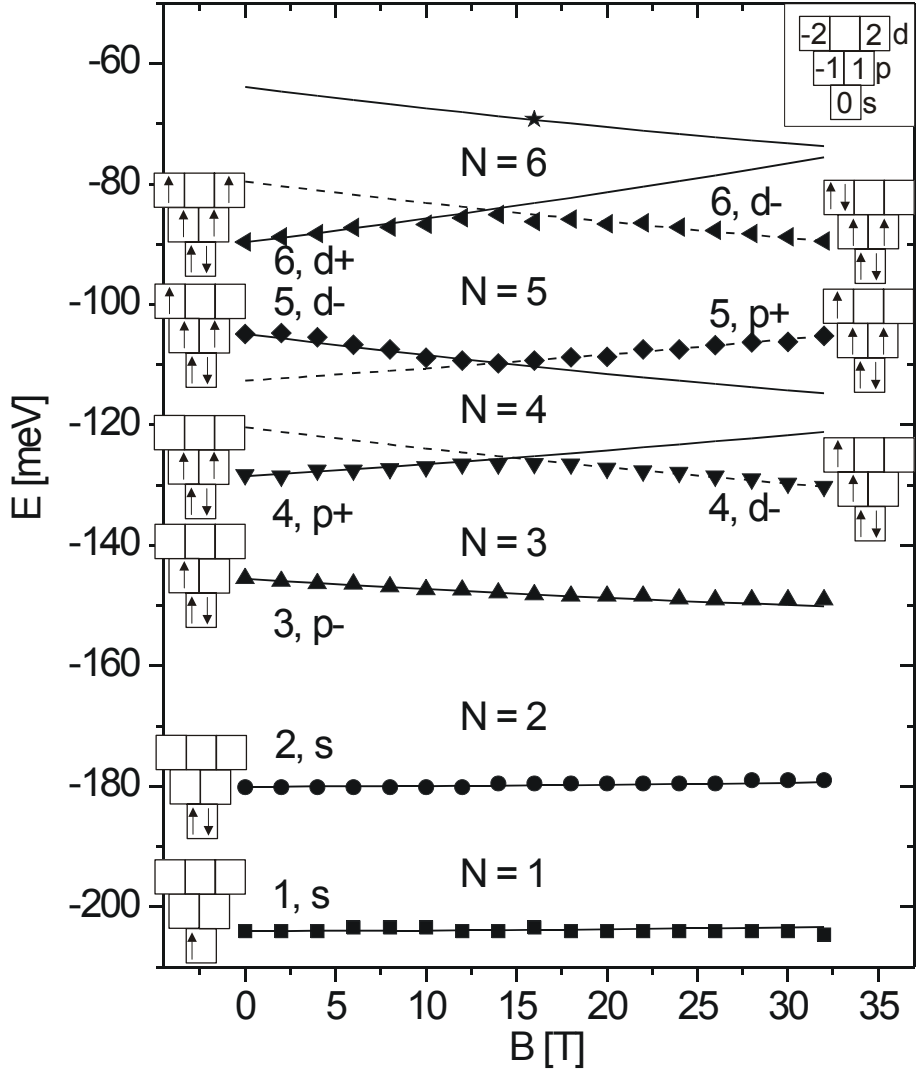

Figure 2: The energetic position of the charging peaks (symbols) from Fig. 1 as function of the magnetic field. The peaks are identified by the same labels as in Fig. 1. N is the total number of holes per QD. The nature of the states with respect to their angular momentum is indicated at the individual graphs. The inserts show the occupation of the single particle levels for each charging peak before and after the level crossing. The arrows represent $J_z = 3/2$ and $J_z = -3/2$, respectively. The assignment of the individual squares is given in the upper left corner of the figure. The solid lines represent fits, using a harmonic oscillator model as outlined in the text.

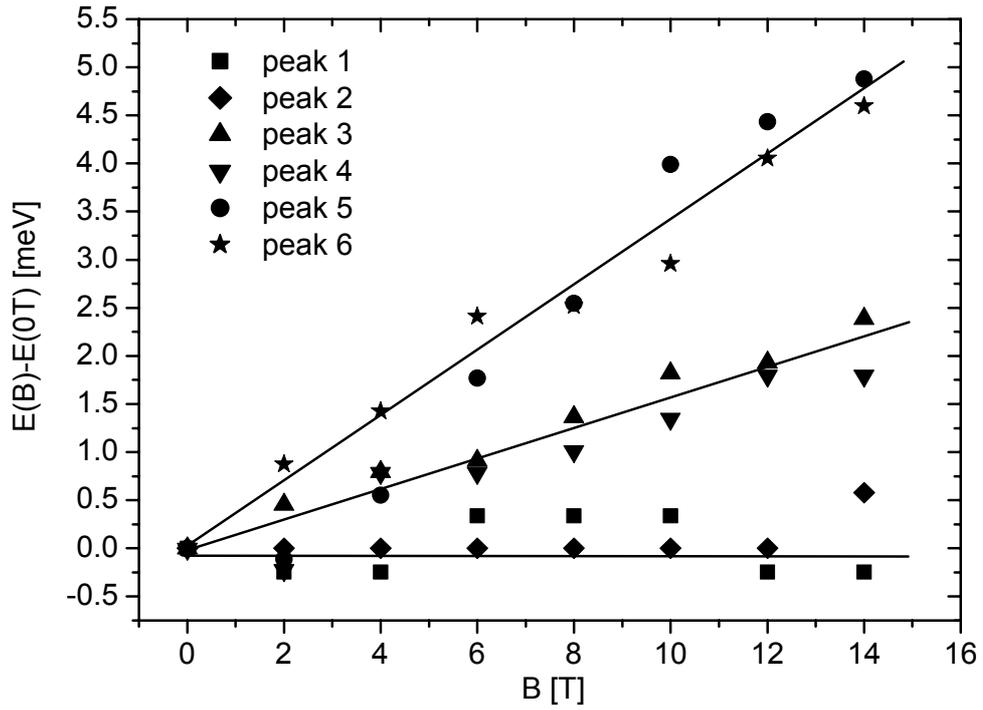

Figure 3: The absolute value of the change in energy |E(B)-E(0T)| of the positions for the charging peaks 1 to 6 (labels according to Fig. 1), shown for fields up to 15 T. Three classes of slopes can clearly distinguished, corresponding angular momenta: Peak 1 and 2 show almost no shift ($L_z = 0$), whereas 5 and 6 have twice the slope ($L_z = \pm 2$) as 3 and 4 ($L_z = \pm 1$).